\documentclass[fleqn,usenatbib]{mnras}
\usepackage[dvipsnames]{xcolor}
\usepackage{newtxtext,newtxmath}
\usepackage[normalem]{ulem}

\usepackage[T1]{fontenc}
\usepackage{ae,aecompl}
\usepackage{times}
\usepackage{graphicx}
\usepackage{amsmath}  

\def\R500c{R_{\rm 500c}}
\def\R200m{R_{\rm 200m}}
\def\M200m{M_{\rm 200m}}
\def\M500c{M_{\rm 500c}}

\newcommand{\change}[1]{{ #1}}

\def\figdir{./figures}

\def\cen{{\rm \, cm}}

\def\Mpc{{\rm \, Mpc}}
\def\kpc{{\rm \, kpc}}
\def\Gyr{{\rm \, Gyr}}

\def\kms{{\rm \, km \, s}^{-1}}

\def\Kelvin{{\rm \, K}}

\def\Msun{M_{\odot}}
\def\shellfish{ \textsc{Shellfish}}

\title[Shock and Splash around $\Lambda$CDM Galaxy Clusters]{Shock and Splash: Gas and Dark Matter Halo Boundaries around $\Lambda$CDM Galaxy Clusters}

\author[Aung, Nagai, \& Lau]{
Han Aung$^{1}$\thanks{E-mail: han.aung@yale.edu} ,
Daisuke Nagai$^{1}$,
Erwin  T.\ Lau$^{1,2}$
\\
$^1${Department of Physics, Yale University, New Haven, CT 06520, U.S.A.}\\
$^2${Department of Physics, University of Miami, Coral Gables, FL 33124, U.S.A.}
}

\pubyear{2020}

\begin{document}
\label{firstpage}
\pagerange{\pageref{firstpage}--\pageref{lastpage}}
\maketitle

\begin{abstract}
Recent advances in simulations and observations of galaxy clusters suggest that there exists a physical outer boundary of massive cluster-size dark matter haloes. In this work, we investigate the locations of the outer boundaries of dark matter and gas around cluster-size dark matter haloes, by analyzing a sample of 65 massive dark matter halos extracted from the {\em Omega500} zoom-in hydrodynamical cosmological simulations. We show that the location of accretion shock is offset from that of the dark matter splashback radius, contrary to the prediction of the self-similar models. The accretion shock radius is larger than all definitions of the splashback radius in the literature by $20\%-100\%$. The accretion shock radius defined using the steepest drop in the entropy and pressure profiles is approximately $1.89$ times larger than the splashback radius defined by the steepest slope in the dark matter density profile, and it is $\approx1.2$ times larger than the edge of the dark matter phase-space structure. We discuss implications of our results for multi-wavelength studies of galaxy clusters.
\end{abstract}

\begin{keywords}
cosmology: theory -- dark matter --  large-scale structure of Universe -- galaxies: clusters: general -- galaxies: groups: general -- methods: numerical 
\end{keywords}

\section{Introduction}
\label{sec:intro}

In recent years, the outskirts of galaxy clusters have emerged as one of the new frontiers for cosmology and astrophysics \citep[see][for review]{walker_etal19}. 
Recent theoretical advances revealed that the physical outer boundary for a dark matter (DM) halo can be defined using by the ``splashback'' radius based on the DM density profile drop \citep[e.g.,][]{diemer_kravtsov2014,adhikari_etal2014,more_etal2015}, the aspherical splashback surface \citep{mansfield_etal17,Mansfield2020}, or the edge radius of the DM phase space structure \citep{Aung2020}, with various definitions encompass varying fraction of orbiting DM particles \citep{diemer_etal17}. Observationally, the outer boundaries of the DM haloes have recently been detected using weak-lensing \citep[e.g.,][]{chang_etal18}, galaxy number density \citep[e.g.,][]{more_etal2016,Baxter2017,Shin2019,Zurcher_More2019,Murata2020} and phase space structure \citep{Tomooka2020}. Upcoming multiwavelength surveys (such as CMB-S4 in microwave and DESI, Rubin, PFS in optical) will provide unprecedented insight on the outer boundaries of massive DM haloes and promise to shed new insight into cosmology and non-linear structure formation of the Universe.

Gas accreting at cluster outskirts provides an alternative probe of cluster boundary. However, the dynamics of the collisional gas is fundamentally different from that of collisionless DM. Unlike the collisionless DM particles which orbit within the DM halos, the collisional gas is shock heated during its first infall, resulting in a high Mach number ($\mathcal{M}>100$) cosmic accretion shock marked by the prominent entropy jump. The secondary infall model predicts that the location of the accretion shock coincides with the splashback radius \citep{bertschinger1985, shi2016b}. Commonly referred to as ``external shock'' in the literature \citep[e.g.,][]{Miniati2000,ryu_etal2003,Skillman2008, molnar_etal2009}, the accretion shock arises from the infall of low density pristine gas in the void regions onto the cluster potential (in contrast to ``internal shocks'' which occurs within the virialization region of DM haloes due to mergers and penetrating filaments). The external accretion shock thus defines a physical boundary of the hot collapsed gas in DM haloes, which is also dependent on their mass accretion rate (MAR) \citep{lau_etal2015}. 

In this work we investigate the locations of shock and splashback radii by analyzing the {\em Omega500} hydrodynamical cosmological simulations. We find that the accretion shock radius defined using the drop in the gas entropy is larger than all definitions of the splashback radius in the literature by $20-100\%$, in contrast to the prediction of the self-similar models. Specifically, we find that the accretion shock radius is larger by $\approx 1.89$ relative to the splashback radius and $\approx 1.2$ relative to the edge radius of the DM phase space structure. Furthermore, we find that the ratios of the shock and splashback/edge radii are independent of halo mass and redshift, but dependent on their MAR. 

We describe our simulations and analysis methods in \S\ref{sec:sim}. Results and discussions are presented in \S\ref{sec:results} and \S\ref{sec:discussions}, respectively. Conclusions are summarized in \S\ref{sec:conclusions}. 

\section{Simulations}
\label{sec:sim}

We analyze the clusters from the {\em Omega500} simulation \citep{nelson_etal2014}, a high-resolution hydrodynamical simulation of a large cosmological volume 
with the comoving box size of $500\,h^{-1}\,\Mpc$. 
The simulation is performed using the Adaptive Refinement Tree (ART) $N$-body+gas-dynamics code \citep{kravtsov1999, kravtsov_etal2002, rudd_etal2008}, 
which is an Eulerian code that uses adaptive refinement in space and time, 
and non-adaptive refinement in mass \citep{klypin_etal2001} 
to achieve the dynamic ranges to resolve the cores of haloes formed in self-consistent cosmological simulations in a flat $\Lambda$CDM model with WMAP 5 years cosmological parameters: 
$\Omega_m = 1 - \Omega_{\Lambda} = 0.27$, $\Omega_b = 0.0469$, 
$h = 0.7$ and $\sigma_8 = 0.82$, where the Hubble constant is defined as 
$100\,h\;\kms\,\Mpc^{-1}$ and $\sigma_8$ is the mass variance within 
spheres of radius $8\,h^{-1}\Mpc$. 

Haloes are identified in the simulation using a spherical overdensity halo finder described in \cite{nelson_etal2014}.  We select 65 haloes with mass $M_{\rm 500c} \geq 3\times10^{14}\,h^{-1}\Msun$ at $z=0$ and re-simulate the box with the higher resolution DM particles in regions of the selected haloes, resulting in an effective mass resolution of $1.09 \times 10^9\, h^{-1}\Msun$, which corresponds to $2048^{3}$ DM particles. We built the merger tree by tracking the most massive progenitors of haloes over time using the merger tree code presented in \citet{yu_etal2015}. This is done by following the $10\%$ most bound DM particles at each snapshot. We define the start of the merging process at the epoch when $R_{\rm 500c}$ of the two haloes start to overlap with each other. Following \citet{diemer_kravtsov2014}, we compute the MAR of DM haloes as
\begin{equation}
\Gamma_{\rm 200m} = \frac{\log_{10}M_{\rm 200m} (z)  - \log_{10}M_{\rm 200m} (z')}{\log_{10}a(z)-\log_{10}a(z')} \; ,
\end{equation}
where $a$ is the expansion factor and $z=0$, $z'=0.5$, and $M_{\rm 200m}$ is the mass enclosed within the radius $\R200m$ such that the density enclosed is 200 times the mean density of the universe, $M_{\rm 200m} = 200\rho_m \times 4\pi \R200m^3/3$.  

The simulation is performed on a uniform $512^3$ grid with 8 levels of mesh refinement, implying a maximum comoving spatial resolution of $3.8\,h^{-1} \kpc$. The spatial resolution is controlled by the density of the cells, and the maximum comoving resolution is only achieved at the centre of the haloes. However, the spatial resolution near the shock radius is between $0.03\,h^{-1}\Mpc$ and $0.12\,h^{-1}\Mpc$, which is sufficient to determine the locations of the edge and shock radii (which is typically of order several Mpc in size) with the accuracy better than 5\%. With a typical number density of $n_e\approx 10^{-3}\cen^{-3}$ and temperature of $T\approx 10^6\Kelvin$, the mean free path of electron is much smaller than the resolution. 

Since the effects of non-gravitational baryonic physics (such as gas cooling and energy feedback from supernova and black holes) are small in cluster outskirts compared to cluster cores, we focus on analyzing the outputs of the non-radiative simulation for simplicity. 
For completeness, we also checked the effects of baryonic physics by comparing the results to those of runs with cooling and star formation and AGN feedback.

\section{Results}
\label{sec:results}

\begin{figure}
\begin{center}
\includegraphics[scale=0.40]{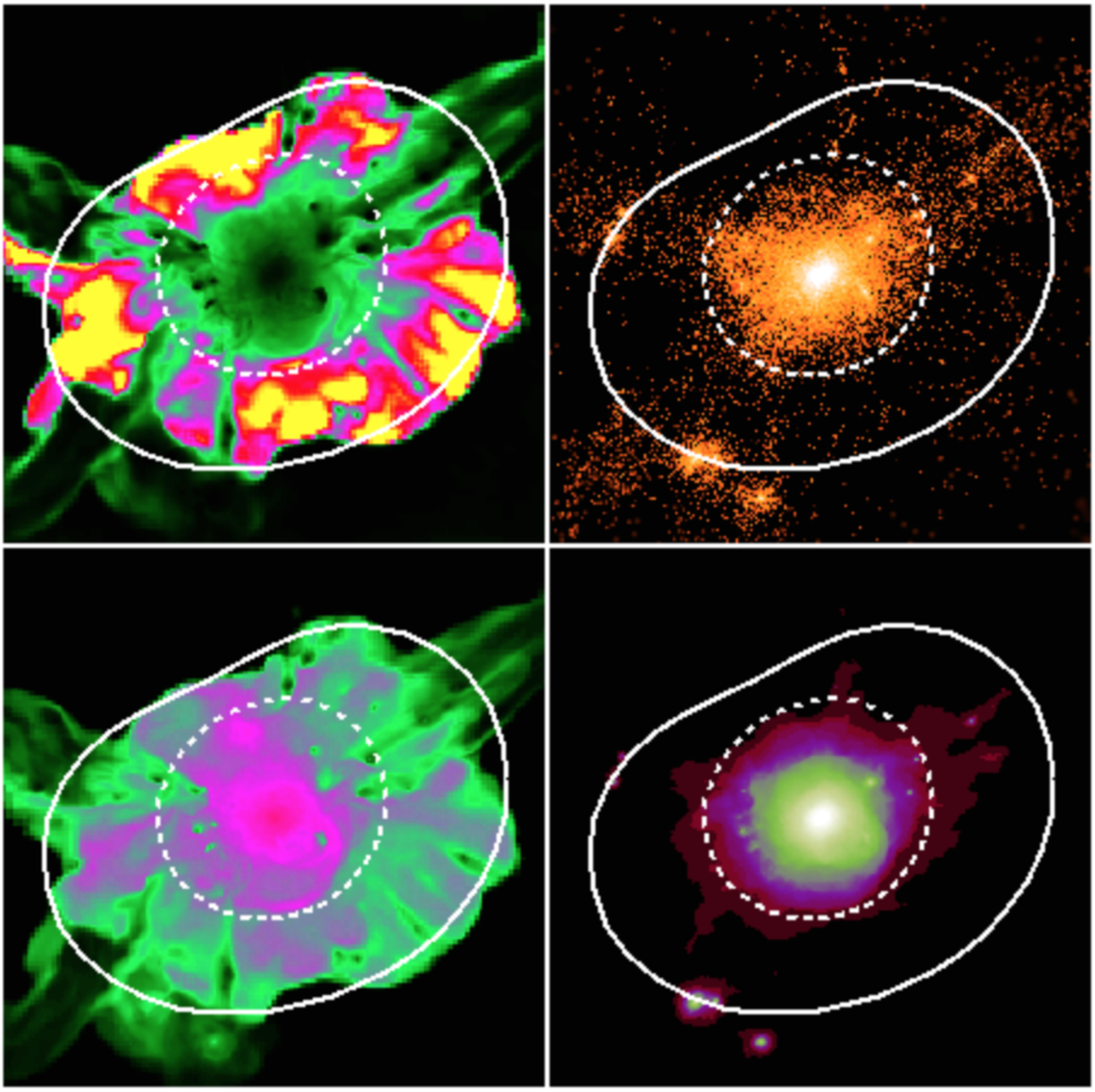}
\caption{Gas entropy (top left), DM density (top right), gas temperature (bottom left) and gas pressure (bottom right) maps of the simulated cluster (CL135) extracted from the non-radiative {\em Omega500} hydrodynamical cosmological simulation. The images are 15.625Mpc/h wide with the projection depth of 3.90625 Mpc/h.  The inner dashed lines indicate the splashback shell computed using the method from \citet{mansfield_etal17}, whereas the outer white lines indicate the shock shell found by the discontinuous jump in entropy as well as pressure. Note that several low-entropy gas streams have penetrated inside of the accretion shock radius along the filaments without getting shock heated.
}
\label{fig:rsp_rsh_map}
\end{center}
\end{figure}

\subsection{Identifying Splashback and Shock Surface}

We determine the location of the splashback radius using the \shellfish\ code \citep{mansfield_etal17}. For each halo, the code draws $10^5$ random sight lines from the halo centre and samples the DM density along each line-of-sight (LOS). The splashback radius $R_{\rm sp}$ is defined as the radius of a spherical surface that encompasses the same volume as that enclosed by the surface of sharp DM density jumps in all LOS \citep[see][for more details]{mansfield_etal17}. 

We determine the accretion shock radius in a similar manner using the \shellfish\ code. Namely, we draw $786$ LOS according to HEALPix \citep{Healpix} pixels from the halo centre and sample the gas entropy profile along each LOS. The 786 LOS chosen here corresponds to the fourth level resolution of HEALPix, and at the shock radius, each pixel corresponds to about $0.26\,h^{-2}\Mpc^2$, with a length resolution of $\approx 0.5\,h^{-1}\Mpc$, about 4 times the simulation resolution in the region. Measuring the shock radius with the fifth level resolution with 3072 LOS leads to a less than $3\%$ difference in shock measurements. 

For each LOS, we select all the gas cells that each LOS passes through, and sample the gas entropy profile along each LOS. The profile is then smoothed with a Savitzky-Golay (SG) filter with window-length of 9 equally spaced logarithmic radial bins and a polynomial order of 5; we checked that the results are robust to the variation in the parameter for SG filter (in the window-length from 5 to 11 bins) and polynomial orders (in the range of 2 to 7). We define the location of the accretion shock as the radius of the minimum in the logarithmic entropy slope. We remove LOS where the entropy jump is less than a factor of 50 (corresponding to Mach number of $\mathcal{M} \sim 20$; the results are unchanged for $\mathcal{M}=10$ to $50$), as these directions host substructures and filamentary gas streams. 
After these removals, the covering fraction of the shock surface of the total spherical area is approximately $80\%$ for all clusters. 
We fit Penna-Dines function with $K=1,I=J=2$ \citep[corresponding to the first and second order expansion of $\cos\phi$ and $\sin\phi$ of polar angle $\phi$,][]{Penna_Dines2007}:
\begin{equation}\label{eq:penna_dines}
    r(\theta,\phi) = \sum_{i,j,k=0}^{I,J,K} c_{\rm ijk} \sin^{i+j}\phi \cos^k\phi \sin^j\theta \cos^i\theta,
\end{equation}
to the shock position along each LOS and calculate the enclosed volume inside the fitted (non-spherical) surface. We then define the shock "radius" as the radius of a sphere that encompasses the same amount of volume as that enclosed within the fitted surface. We note that the total volume inside the shock radius is larger than the total volume of shock heated gas as it also includes volume of unshocked gas residing in filaments.

Figure~\ref{fig:rsp_rsh_map} shows the map of one of the haloes indicating the splashback and shock shells identified using the above algorithms. The splashback shell identified encompasses the DM structure, whereas the shock shell encompasses a much larger, extended area, where the entropy and pressure shows significant decline. We also note that there are several unshocked, low-entropy gas streams that have penetrated inside of the accretion shock radius of the halo along the filaments.

\subsection{Phase Space Structures of DM and Gas}\label{sec:phase_space}

\label{sec:dynamics}
\begin{figure}
\begin{center}
\includegraphics[width=0.49\textwidth,trim={10 1 0 1},clip]{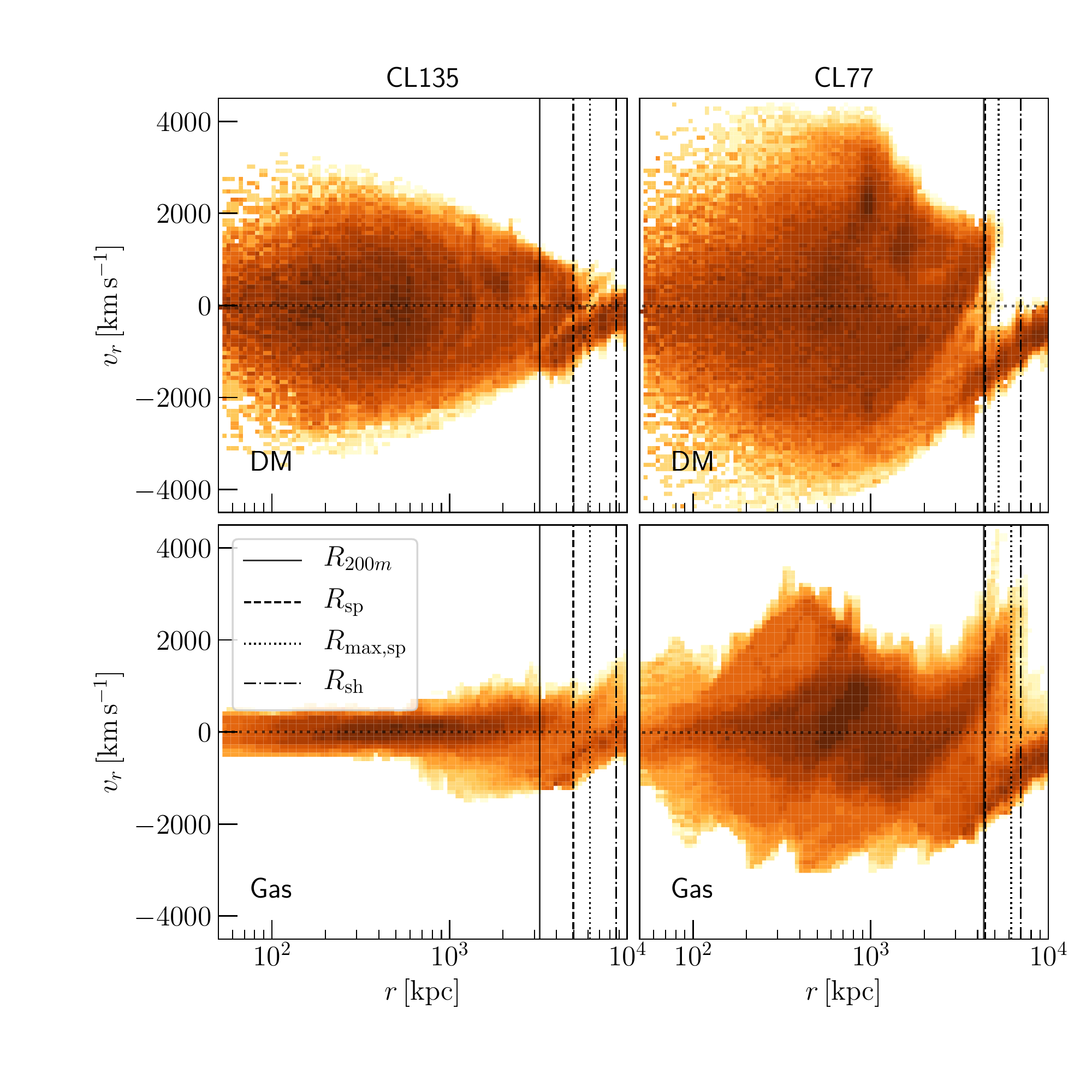}
\caption {Phase-space diagrams of DM ({\em top} panels) and gas ({\em bottom} panels) for a relaxed cluster CL135 ({\em left} panels) with low MAR ($\Gamma = 0.5$), and a merging cluster CL77 ({\em right} panels) with high MAR ($\Gamma = 2.9$).  The solid, dashed, dotted, and dot-dashed lines represent $\R200m$, $R_{\rm sp}$ (splashback radius), $R_{\rm max,sp}$ (maximum splashback radius along any line-of-sight), and $R_{\rm sh}$ (shock radius), respectively. The colour represents the relative mass fraction of DM (top panels) and gas (bottom panels), with deeper colour indicating higher mass fraction at a given radius. The phase space structure of virialized DM haloes extends past $R_{\rm sp}$, reaching nearly $R_{\rm sh}$. Gas follows DM at $r\gtrsim R_{\rm sh}$, while gas dynamics differs significantly from that of collisionless DM at $r\lesssim R_{\rm sh}$.}
\label{fig:phase_diagram}
\end{center}
\end{figure}

Figure~\ref{fig:phase_diagram} shows phase-space densities of DM and gas for their radial velocity components as a function of radius, for a relaxed cluster (CL135) and a merging cluster (CL77), respectively. The average radial velocity of gas and DM is negative outside the shock radius ($R_{\rm sh}$) as they fall onto the cluster potential. The two trace each other as the gas pressure is low, rendering gas to behave similarly to collisionless DM. Gas infalling from the void is shock heated at the shock radius, causing gas to lose its kinetic energy into heating of the gas. Filamentary accretion can bring infalling gas inside of the virialization region of the DM haloes and form shocks with small Mach number ($\mathcal{M}<3$) before gas loses their kinetic energy. The DM particles, on the other hand, exchange energy through gravitational interactions as they orbit through the interior of the DM halo.
Within the splashback radius ($R_{\rm sp}$), a DM halo exhibits a typical virialized phase-space structure, where the splashback and orbital motions of DM particles produce a dispersion with zero mean radial velocity. The phase space structure of the DM halo can extend out to radii larger than the volume-averaged splashback radius as not all particles within the splashback surface are expected to lie within the volume-averaged splashback radius $R_{\rm sp}$ due to asphericity \citep{mansfield_etal17,diemer_etal17,Aung2020}. Even when using $R_{\rm max,sp}$ (the maximum radius of the splashback surface defined using the Penna-Dines approximation surface), a small amount (upto $1\%$) of orbiting particles can still exist beyond $R_{\rm max,sp}$ \citep{Mansfield2020} as shown in Figure~\ref{fig:phase_diagram}.

Due to the collisional nature of the gas, however, the phase-space distribution of gas differs significantly within the interior of the cluster. The radial velocity dispersion of gas is considerably smaller because, as the gas is shock heated through the accretion shock, where most of the gas kinetic energy is converted to thermal energy. Thus, inside $\R200m$, the level of gas motions inside the accretion shock remains small in absence of external disruption by mergers for CL135, while gas motions induced by mergers comprise of most of the velocity dispersion in CL77. The position of the shock radius is closer to the edge of DM phase space than the splashback radius.

\subsection{Shock Radii Determined from Profiles}

\begin{figure}
\begin{center}
\includegraphics[width=0.5\textwidth]{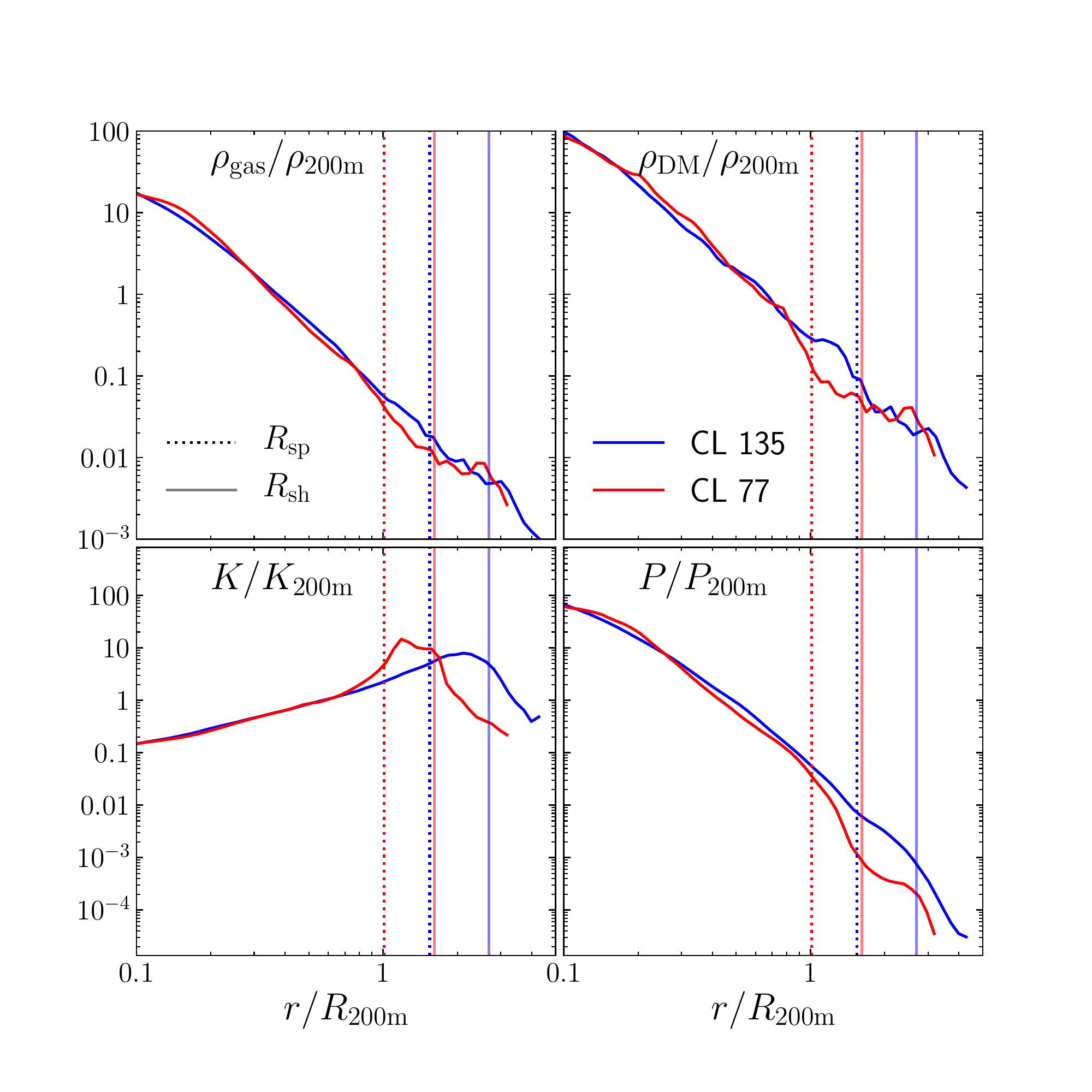}
\caption{Spherically averaged gas density, DM density, volume-weighed entropy and pressure as a function of radius for the two different clusters: a relaxed cluster CL135 with low MAR ($\Gamma = 0.5$), and a merging cluster CL77 ({\em right} panels) with high MAR ($\Gamma = 2.9$). The vertical lines indicate splashback and shock radii based on the \shellfish\ and our method, respectively. The gas pressure and entropy profiles show significant decrease near the shock radii, while the gas density and DM density decreases at the splashback radii. The gas density slope is shallower than the DM density slope. The slower accreting halo (CL135) also has a smoother jump and larger shock and splashback radii than the fast accreting CL77.}
\label{fig:profiles}
\end{center}
\end{figure}

\begin{figure}
\begin{center}
\includegraphics[width=0.49\textwidth]{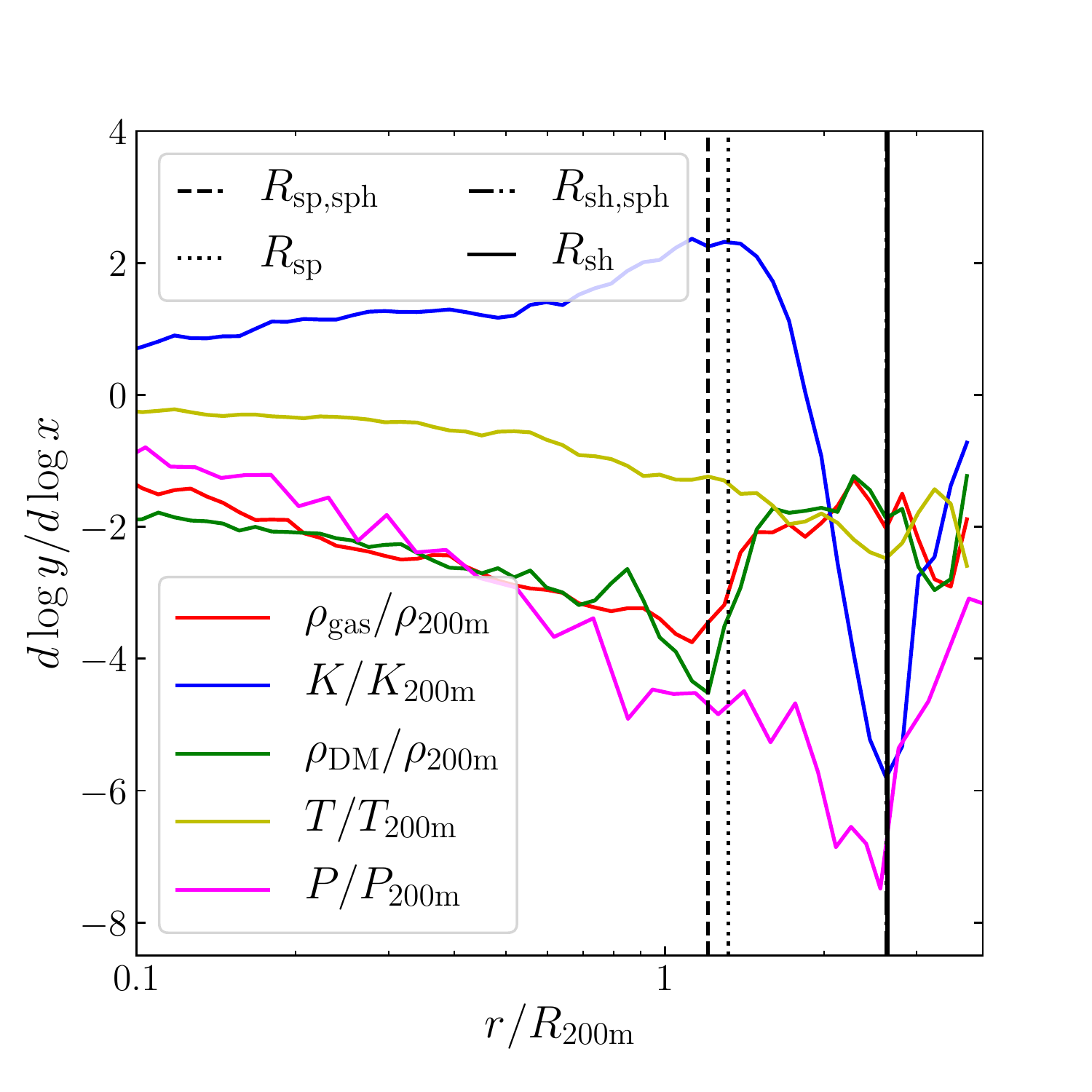}
\caption{The logarithmic slope of different gas and DM median profiles as a function of radius. The radius where the minimum of gas entropy, temperature and pressure slope is defined as the shock radius $R_{\rm sh,sph}$ of the spherically averaged gas profile, denoted by dotted line. The radius where the minimum of DM density slope is defined as the splashback radius $R_{\rm sp,sph}$ of the spherically averaged profile, denoted by dashed line, and coincides with the minimum of the gas density slope.  }
\label{fig:profiles_slope}
\end{center}
\end{figure}

Figure~\ref{fig:profiles} shows the spherically averaged DM density, gas density, and volume-weighed entropy and pressure profiles for two representative clusters in the sample, CL135 a relaxed cluster with low $\Gamma$, and CL77, a merging cluster with high $\Gamma$. We also overplot their splashback radii estimated from \shellfish\ $R_{\rm sp}$, and their shock radii from the shock surface $R_{\rm sh}$. Both $R_{\rm sp}$ and $R_{\rm sh}$ are smaller for the high $\Gamma$ cluster CL77 than than the low $\Gamma$ cluster CL135. In both clusters, there are sharp drops in the pressure and entropy profiles at the accretion shock, and they are particularly prominent in the high $\Gamma$ cluster CL77. However, the corresponding decrease in gas density profiles are small at the accretion shock. This is because the density contrast across a shock is intrinsically smaller than their counterparts in pressure and entropy, and is capped at a maximum value of $4$, as expected from the Rankine -- Hugoniot shock jump condition.

Figure~\ref{fig:profiles_slope} shows the logarithmic slope of the spherically averaged median DM density, gas density, temperature, pressure, and entropy profiles for all clusters at $z=0$. The entropy profile is increasing at all radii before the shock radius, and within $0.1\leq r/\R200m\leq 0.6$, it is consistent with previous findings of entropy slope of $1.1$ \citep[e.g.,][]{voit_etal2005} as expected from self-similar cluster growth. At the shock radius, entropy decreases sharply indicating a strong shock. The shock front, however, is wider and is not as abrupt as the shock front in LOS profile due to smoothing over aspherical shock fronts and variations among clusters. The pressure profile has the minimum slope about $-7$ at the shock radius. Inside the shock radius, the pressure profile is also rapidly decreasing, consistent with the universal pressure profile characterized by the generalized NFW profile \citep[e.g.,][]{nagai_etal2007a,arnaud_etal2010}. Similarly, the temperature profile has the minimum slope of about $-2$ at the shock radius. 

The DM density slope follows  NFW profile \citep{nfw1996} closely, where the slope is $-1$ in the inner region and slowly decreases to $-3$ in outer region before hitting minimum at the splashback radius. The gas density is much flatter in inner region starting with slope of $\approx 0$, but approaches NFW and follows DM profile at outer radii. In fact, gas density slope becomes minimum at DM splashback radius, while only showing mild decrease at the shock radius. The smaller decrease in density slope at the shock radius compared to other thermodynamic quantities is expected, as the density jump for a shock wave in ideal gas is capped at $4$, while there is no upper bound for the jumps in temperature, pressure or entropy. 

For DM, the splashback radius identified as the steepest point in the spherically averaged density profile $R_{\rm sp, sph}$ is smaller than the splashback radius estimated from \shellfish\ $R_{\rm sp}$ \citep[see also][]{mansfield_etal17}. 
For the accretion shock radius, the radius computed from spherically averaged profile $R_{\rm sh, sph}$, is the same as that identified from the volume-averaged shock surface $R_{\rm sh}$. 

\begin{figure}
\begin{center}
\includegraphics[width=0.51\textwidth]{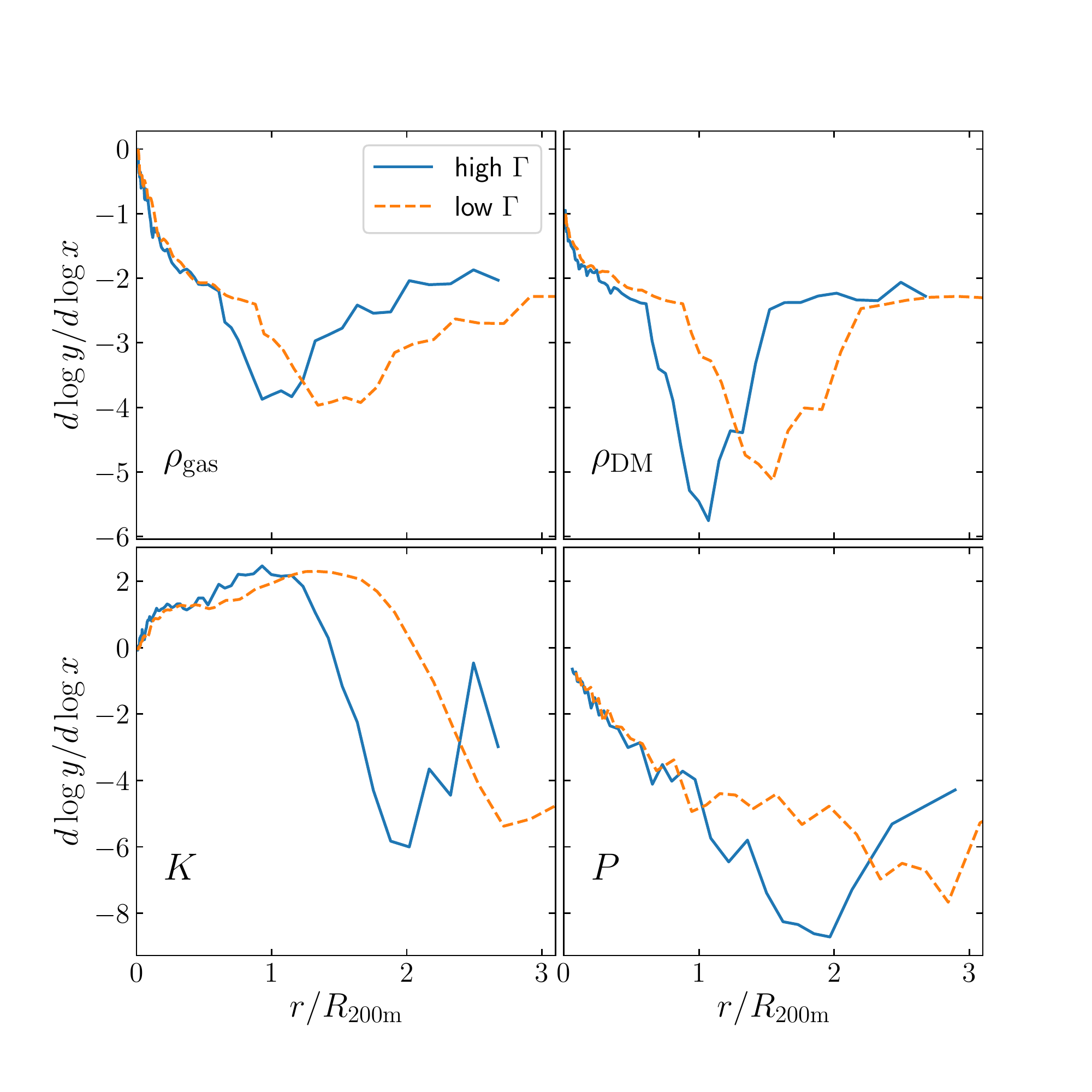}
\caption{The logarithmic slope of different gas and DM median profiles for different MAR. The MAR is split at the 33 and 66-percentile of all haloes, which results in the low $\Gamma_{\rm 200m}$ sample with $\Gamma_{\rm 200m}<1.5$ (dashed line) and high $\Gamma_{\rm 200m}$ sample with $\Gamma_{\rm 200m}>2.7$ (solid line). The splashback radius is where the total matter (DM+gas) density slopes is the smallest, and the shock radius where the minimum of entropy and pressure slope is smaller for the higher MAR haloes. }
\label{fig:mar_slope}
\end{center}
\end{figure}

Figure~\ref{fig:mar_slope} shows the profile slopes for different $\Gamma$ samples. Here the splashback and shock radii are identified as the steepest jumps in DM density and gas entropy profiles, respectively. Both radii decrease for larger $\Gamma$. The DM density slope is steeper for larger $\Gamma$, consistent with the previous result \citep{diemer_kravtsov2014,more_etal2015}. The pressure jump is larger for larger $\Gamma$, indicating a stronger shock. 

\subsection{Offsets between Splashback and Shock Radii}

\begin{figure}
\begin{center}
\includegraphics[width=0.49\textwidth]{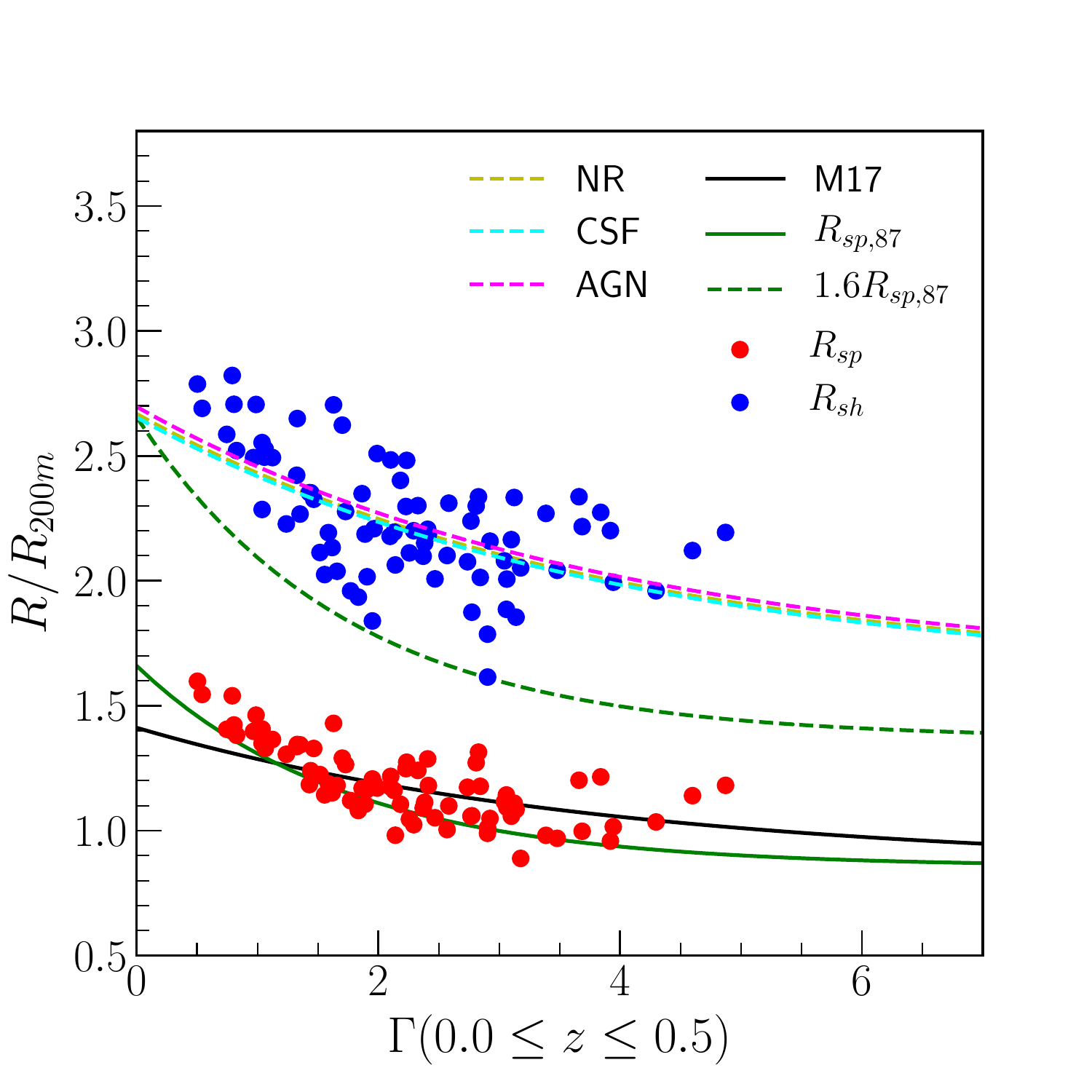}
\caption{The splashback radius $R_{\rm sp}$ (red points) and the accretion shock radius $R_{\rm sh}$ (blue points), normalized by the halo radius $\R200m$, plotted as a function of the MAR of the cluster-size DM haloes, $\Gamma_{\rm 200m}$. 
The solid line is the $R_{\rm sp}-\Gamma_{\rm 200m}$ relation from DM-only cosmological simulation defined by \shellfish\ \citep[M17,][]{mansfield_etal17} and $R_{\rm sp,87}$ computing using SPARTA \citep{diemer_etal17}. The dashed lines represent the best-fitting $R_{\rm sp}-\Gamma_{\rm 200m}$ relation times the average $R_{\rm sp}/R_{\rm sp}$ ratios for three different baryonic simulations (magenta for AGN feedback simulation, cyan for cooling and star forming simulation, and yellow for non-radiative, embedded behind cyan line). Also shown is $r_{\rm edge}\approx 1.6R_{\rm sp,87}$, marking the edge of the DM phase space structure \citep{Aung2020}. 
}
\label{fig:rsp_rsh}
\end{center}
\end{figure}

Figure~\ref{fig:rsp_rsh} shows the splashback radius $R_{\rm sp}$ normalized by the halo radius $\R200m$, plotted as a function of the MAR ($\Gamma_{\rm 200m}$) of haloes extracted from the non-radiative {\it Omega500} simulation.  The splashback radius decreases with increasing MAR, confirming previous numerical \citep{diemer_kravtsov2014,mansfield_etal17} and analytic results \citep{adhikari_etal2014, shi2016a}. 
The $R_{\rm sp}$ from our hydrodynamic simulation agrees well with the best-fitting relation from the DM-only simulation. 

In the same panel, we show that $R_{\rm sh}/\R200m$ decreases with $\Gamma_{\rm 200m}$ in a similar manner to the $R_{\rm sp}$--$\Gamma_{\rm 200m}$ relation. The average ratio between the radii is $R_{\rm sh}/R_{\rm sp} = 1.89 \pm 0.16$ (based on the yellow dashed line in Figure~\ref{fig:rsp_rsh}, where the error indicates 1$\sigma$ scatter) at $z=0$, and it is only weakly dependent on MAR for the range probed here. At $z=1$, $R_{\rm sh}/R_{\rm sp} = 2.03 \pm 0.32$, and at $z=3$, $R_{\rm sh}/R_{\rm sp} = 2.12 \pm 0.35$ which is consistent with no evolution with redshift. We also find that the scatter in $R_{\rm sh}$ is larger than that of $R_{\rm sp}$. 

We repeated the same analysis for the same sets of haloes with baryonic physics that include radiative cooling and star formation, and AGN feedback. The differences in $R_{\rm sp}$ and $R_{\rm sh}$ between simulations with different baryonic physics is $\lesssim 1$\%, thus both radii remain essentially unchanged in the presence of baryonic physics. 

We note that the phase space structure of the DM halo can extend out to radii larger than the splashback radius, because not all particles within the splashback surface are expected to be enclosed within the volume-averaged $R_{\rm sp}$. In fact, the splashback radius from \shellfish\ only contains about $87\%$ of the particle apocentre \citep{diemer_etal17}. The edge radius, which marks the end of DM phase space, corresponds to the radius where the fraction of orbiting subhaloes is greater than 99\% (denoted as $R_{\rm sp,99}$), which is approximately $1.6$ times larger than the splashback radius measured with \shellfish\ (denoted as $R_{\rm sp,87}$) \citep{Aung2020}. This edge radius lies in the region in between the accretion shock and splashback surfaces. Specifically, the ratio of the shock and edge radius is $R_{\rm sh}/1.6R_{\rm sp,87} \approx 1.2$ for $1\leq \Gamma_{\rm 200m} \leq 4$, indicating that the shock radius is about 20$\%$ larger than the edge radius, on average.


\subsection{Shapes of Shock and Splashback Shells}\label{sec:apshericity}

\begin{figure}
\begin{center}
\includegraphics[width=0.49\textwidth]{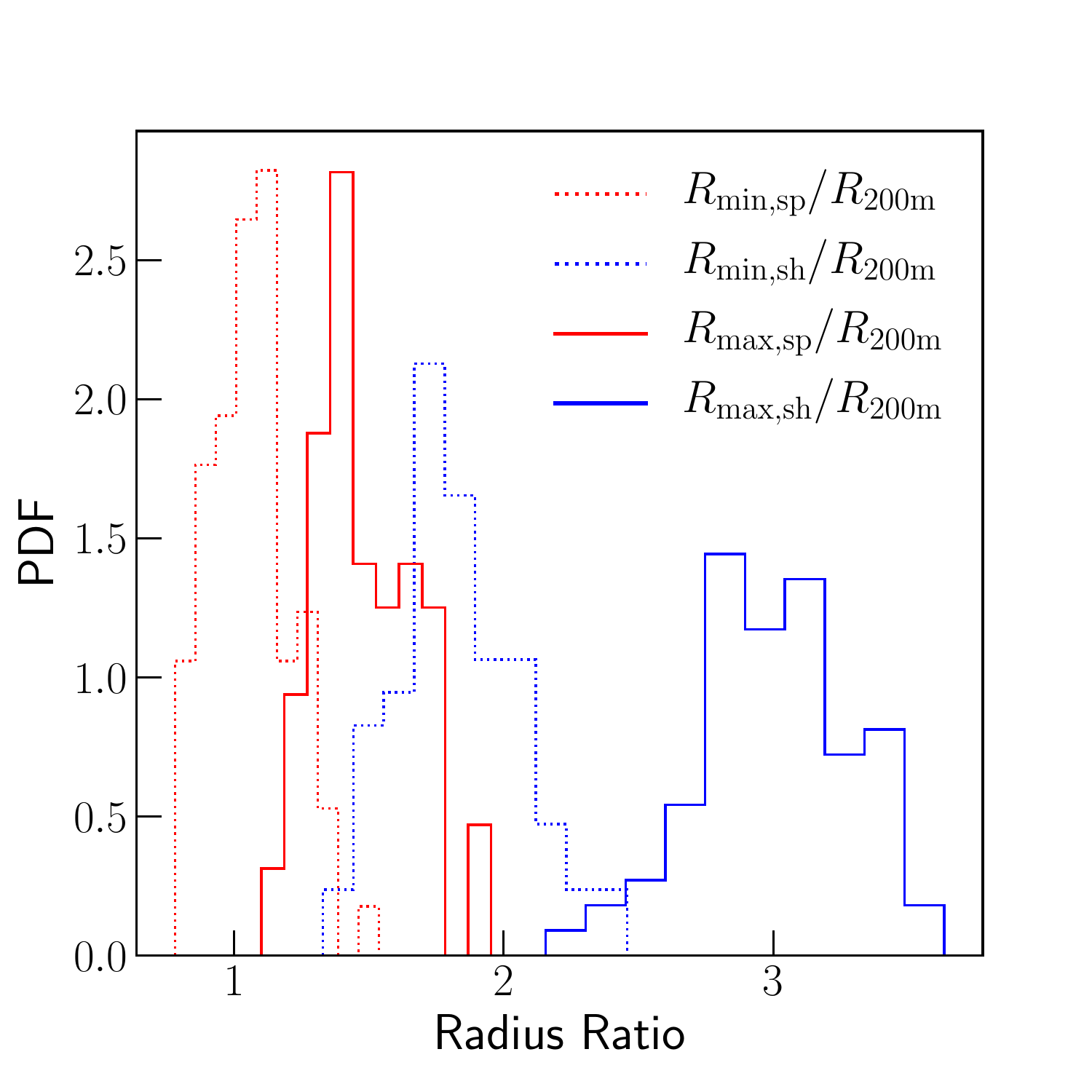}
\caption{\change{The normalized distribution of maximum (solid) and minimum (dotted) radius of shock (blue) and splashback (red) surface with respect to $R_{\rm 200m}$. The shock and splashback shells are highly aspherical, causing the maximum and minimum radius along lines-of-sight to differ by more than $65\%$. Note that the overlap in the distributions of the minimum shock and maximum splashback radii is due to the scatter in $R_{\rm sh}/R_{\rm sp}$ ratio. For individual haloes, the shock radius is always larger than the splashback radius along every line-of-sight.
}}
\label{fig:rmaxmin}
\end{center}
\end{figure}

\begin{figure}
\begin{center}
\includegraphics[scale=0.5]{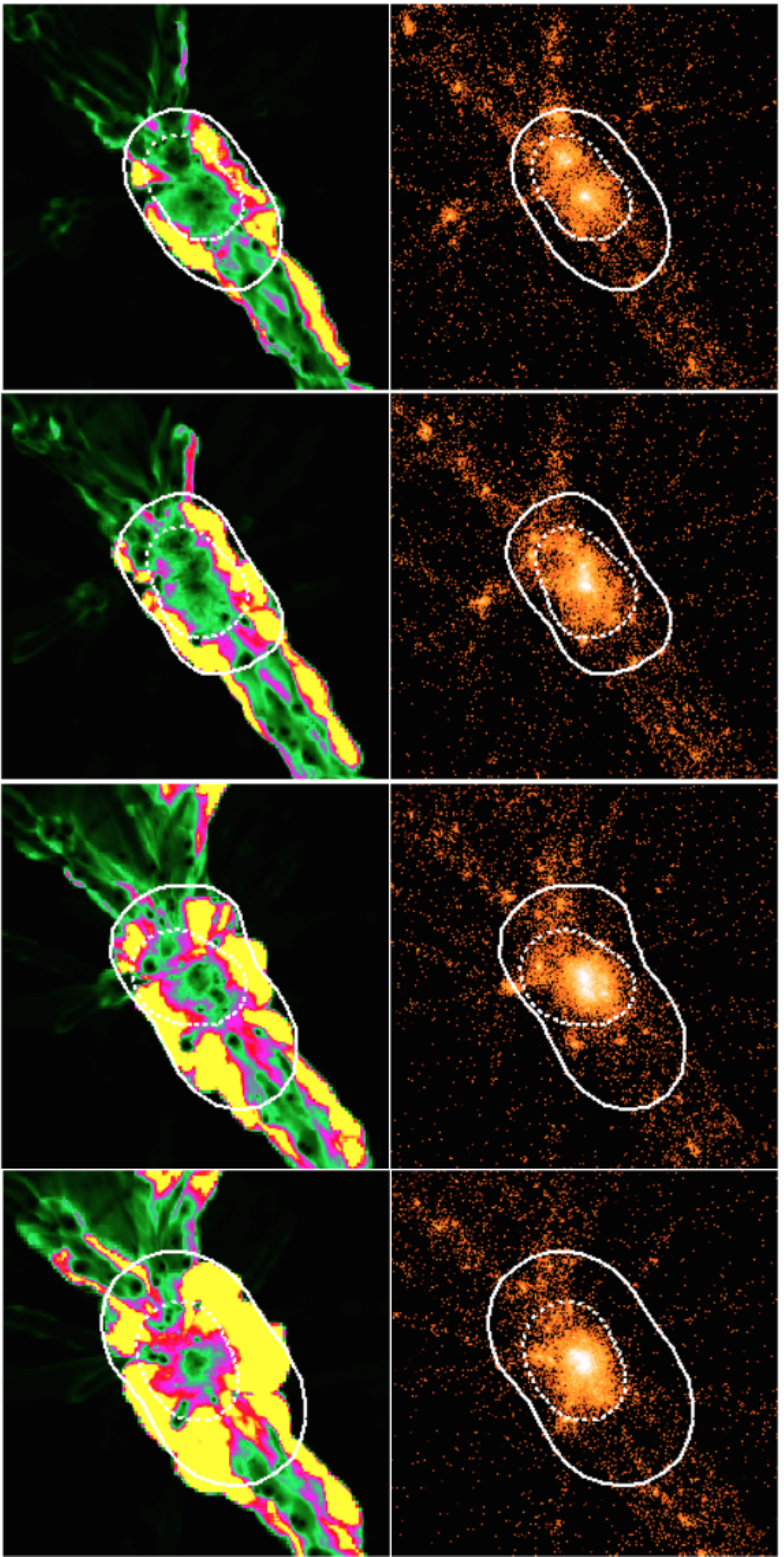}
\caption {Maps of gas entropy ({\em left} panels) and DM ({\em right} panels) of a cluster (CL21) undergoing an almost equal mass (mass ratio = 0.83) merger. The solid line and the dotted line show the accretion shock and splashback shells respectively. 
The panels from {\em top} to {\em bottom} show the cluster at different merging stages: $t_{\rm merge}= -0.4\Gyr, +0.4\Gyr, 
+1.25\Gyr, +2\Gyr$, where $t_{\rm merge}$ is the merging time defined as when $R_{\rm 500c}$ of the two merging haloes first touches. During the merger, the splashback and accretion shock shells continue to evolve. After $2 \Gyr$, the DM splashback shell becomes more spherical, while the accretion shock is still elongated along the axis of merger and filament. 
}
\label{fig:rsh_rsp_merge_map}
\end{center}
\end{figure}

In practice, DM splashback and accretion shock are aspherical, because haloes form through merger and accretion of materials through cosmic web of filaments that are inherently aspherical, as shown in Figure~\ref{fig:rsp_rsh_map}. Figure~\ref{fig:rmaxmin} shows the distribution of the maximum ($R_{\rm max}$) and minimum ($R_{\rm min}$) splashback and shock shells. We find that the $R_{\rm max,sh}$ and $R_{\rm min,sh}$ of the shock shell can be $1.65$ times larger and $0.7$ times smaller compared to the volume-averaged shock radius, respectively, while the maximum and minimum distance of the splashback shell ranges between $(0.8-1.42)$ times the volume-averaged splashback radius. This shows that the accretion shock shells are generally more aspherical than the splashback shells, which cause the spherically averaged gas profiles to appear smoother than the actual accretion shocks. \change{Even though the minimum shock radii and maximum splashback radii overlap in Figure~\ref{fig:rmaxmin}, we emphasize that the largest radius of the splashback shell is still smaller than the smallest radius of the shock shell of the same halo for the individual haloes in our sample. The minimum shock to splashback radius ratio is found in the directions perpendicular to the axis of filament and merger (see the panels in the second and third rows of Figure~ \ref{fig:rsh_rsp_merge_map} where the ratio is as small as $1.06$). }

Major mergers are also responsible for the aspherical shapes in the accretion shock and splashback shells. Even though mergers are not directly responsible for the formation of the accretion shock, the `run-away' shocks generated from mergers \citep{Zhang2019} can overtake and power the accretion shock, thus affecting the shape of the accretion shock more than the shape of the DM splashback shell (see Section~\ref{sec:mergers} for the impact of mergers on accretion shock radius).
In Figure~\ref{fig:rsh_rsp_merge_map}, we show the evolution of gas entropy (the most apparent feature in the accretion shock as seen in Figure~\ref{fig:rsp_rsh_map}) and DM density maps as the cluster undergoes a major merger.  Before the merger, the splashback shell encompasses the two merging sub-clusters, while the accretion shock encloses the shock-heated gas associated with these two subclusters. After the merger, the splashback shell decreases rapidly as the collisionless DM of the two clusters overlap with each other. The accretion shock radius, on the other hand, decreases more slowly with time as the gas lags behind (middle two panels in Figure~\ref{fig:rsh_rsp_merge_map}). Thus, towards the end of the relaxation period, the ratio of shock to splashback radii is slightly larger than that before the merger, $R_{\rm sh}/R_{\rm sp}$ by $\approx 10\%$ (bottom panel in Figure~\ref{fig:rsh_rsp_merge_map}). 
Note that the transient internal shocks driven by mergers have much lower Mach number compared to the external accretion shock, leading to much smaller entropy jumps compared to those produced by the accretion shocks.

\begin{figure}
\begin{center}
\includegraphics[width=0.49\textwidth]{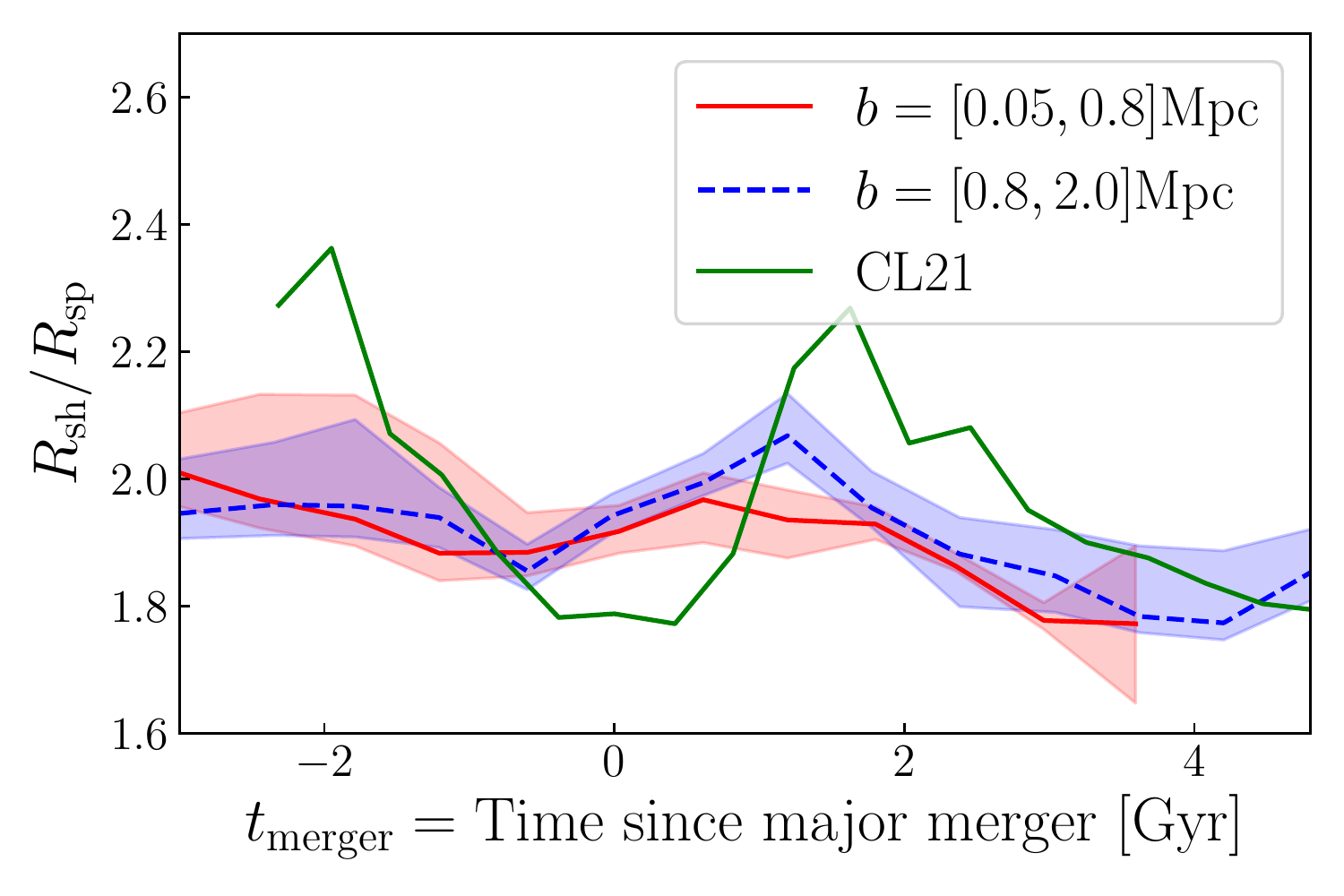}
\caption{The shock to splashback radius ratio as a function of time since major merger for different impact parameters, indicated with solid and dashed lines. The lines and shaded regions indicate median and the 16-84 percentile of the ratio. The green line shows the most extreme major merger in CL21, whose gas entropy and DM density maps are shown in Figure~\ref{fig:rsh_rsp_merge_map}. The $R_{\rm sh}/R_{\rm sp}$ ratios increase temporarily within $1-2\Gyr$ after major mergers, and they all stay within the range of 1.6 and 2.4 throughout the mergers. 
} 
\label{fig:rsh_rsp_merger}
\end{center}
\end{figure}

\subsection{Effects of Mergers on Accretion Shock Radius}\label{sec:mergers}
Mergers can increase the accretion shock radius, boosting the ratio of the shock to splashback radii ($R_{\rm sh}/R_{\rm sp}$).
A merger event can generate a merger shock in the ICM, which accelerates to outer radii through a ``run-away'' shock with high Mach number \citep{Zhang2019}. When a merger shock runs into the accretion shock, the merger shock accelerates the accretion shock and pushes it out temporarily, whose duration depends on the MAR of the halo and the Mach number $\mathcal{M}$ of the shock \citep{Zhang2020}. Higher MAR leads to smaller increase in the accretion shock radius. For example, for a merger shock with $\mathcal{M} = 1.5$, in a halo with low MAR of $\Gamma_{\rm 200m}=1$, the accretion shock radius can grow to twice as predicted by self-similar model, before receding back to self-similar predicted value $3$~Gyr later. However, for the same Mach number merger shock in a halo with a higher MAR of $\Gamma_{\rm 200m}=3$, the accretion shock radius can only grow to 1.5 times the self-similar predicted shock radius, before receding back to the self-similar value after $1.5$~Gyr. Physically, the higher MAR of the halo implies larger ram pressure of the infalling gas, which pushes the run-away shock farther back towards the cluster center, leading to the smaller accretion shock radius.

Figure~\ref{fig:rsh_rsp_merger} shows $R_{\rm sh}/R_{\rm sp}$ before and after major merger events from $z=4$ onward for different impact parameters in the simulation, where a major merger is defined as the merger event where the mass ratio is less than 1:4. Impact parameter $b$ is defined as the perpendicular distance between the paths of the mergers, where the direction of the path is determined by the velocities of the haloes in the last snapshot when $R_{\rm 500c}$ of two haloes do not overlap.
It shows a temporary increase in $R_{\rm sh}/R_{\rm sp}$ right after the merger peaking after 1~Gyr and lasting around 2~Gyr. 
The average $R_{\rm sh}/R_{\rm sp}$ before and after merger remains relatively constant, but can increase by $10\%$ within $2-3\Gyr$ after the merger.  In extreme cases where there is equal mass merger and almost head-on collision (with the mass ratio of $0.83$ and impact parameter $b=0.49$~Mpc for CL21), we find that after 2 Gyr of the merger the $R_{\rm sh}/R_{\rm sp}$ ratio temporarily increases up to 2.4, corresponding to an increase of about $30\%$, before receding back to 1.8 after 4 Gyr of the merger. In fact, the ratio ranges between 1.6 to 2.4 during mergers of all haloes in our sample. 

Our results indicate that mergers cannot account for all of the offsets between the volume-averaged $R_{\rm sh}$ and $R_{\rm sp}$, which is almost a factor of 2. When undergoing major merger, the ratio of volume-averaged radii $R_{\rm sh}/R_{\rm sp}$ of CL21 remains above 1.6 before and after the merger (Figure~\ref{fig:rsh_rsp_merger}). However, in Figure~\ref{fig:rsh_rsp_merge_map}, we found the instance where the shock to splashback ratio becomes very close to 1 and rapidly increases as the merger shock propagates away from the merger axis. This indicates that whether the merger-accelerated shocks can explain the offset between the shock and splashback radii depends on which line-of-sight we choose to compare. We leave further exploration of this idea to a future work.

\section{Discussions \& Future Work}
\label{sec:discussions}

The present work focuses on the massive DM haloes where the accretion shock is growing and the cluster outskirt is not affected significantly by galaxy formation physics. However, the accretion shocks are expected to behave differently for lower-mass group and galaxy scales. Radiative cooling can significantly reduce the pressure support of the galactic haloes which can lead to the collapse of the accretion shock into the inner region \citep{Birnboim03}. In addition, at lower halo masses of $M<10^{12}M_{\odot}$, the DM phase space structure of the orbiting halo can extend out to turnaround radii \citep{Prada2006}, which leads to much larger ratio of $R_{\rm edge/}R_{\rm 200c}\approx4$ \citep{Ludlow2009,Wang2009}. Future studies should extend the current work to lower mass haloes. Additionally, higher mass and spatial resolution simulations will be required to resolve the shock radius of these low mass haloes.

\section{Conclusions}
\label{sec:conclusions}

In this work, we investigate the relation between the splashback of dark matter (DM) and the accretion shock of gas in the outskirts of cluster-size DM haloes, using the {\it Omega500} cosmological hydrodynamical simulations. Our main findings are summarized below: 

\begin{enumerate}
\item The accretion shock radius is located farther from the cluster centre than the DM splashback radius (Figure~\ref{fig:rsp_rsh_map}). The phase space structures of DM and gas follow each other outside the accretion shock where the gas thermal energy is small compared to its kinetic energy. Inside the accretion shock, gas is thermalized with relatively small radial infall velocities, while DM particles orbit with large velocity dispersion within the interior of DM haloes. The phase space structures for both the orbiting DM and the thermalized gas extend beyond DM splashback radius (Figure~\ref{fig:phase_diagram}). 

\item The ratio between the two radii depends on the definitions of the splashback and the shock radii. 
Specifically, the accretion shock radius defined by the entropy drop is larger than all definitions of the splashback radius in the literature.  The accretion shock radius defined using the steepest drop in entropy is approximately $\approx 1.89$ times larger than the splashback radius defined by the steepest slope in the DM density profile, and it is $\approx1.2$ times larger than the edge of the DM phase-space structure (Figure~\ref{fig:rsp_rsh}).

\item The accretion shock radius of gas decreases with mass accretion rate (MAR) of the halo, similar to the MAR dependence in the splashback radius (Figures~\ref{fig:profiles} and \ref{fig:mar_slope}). The resulting ratio is fairly independent of redshift and baryonic physics for the cluster-size DM haloes (Figure~\ref{fig:rsp_rsh}).

\item The gas density follows DM density at the outer radii with steepest density decrease near the DM splashback radius. The steepest decrease in other thermodynamic (temperature, pressure, and entropy) profiles, however, occurs at the location of the accretion shock (Figure~\ref{fig:profiles_slope}). We do not find significant gas density jump in radial profiles at the accretion shock as the ratio of the post- and pre-shock density is capped at a maximum value of $4$ in the strong shock limit, which is further smeared out by azimuthally averaging around aspherical shocks.

\item Mergers account for on average $\approx 10\%$, and upto $30\%$ increase in the shock to splashback ratio for the duration of $\approx 1$~Gyr right after the merger. After that, the merger-accelerated shock recedes back to the pre-merger accretion shock  (Figure~\ref{fig:rsh_rsp_merger}). 
\end{enumerate}

Our results have broad implications for using galaxy clusters as probes of astrophysics and cosmology. First, the apparent spatial offset between splashback and the more extended shock radii indicates that the transition boundaries between the 1-halo and 2-halo terms are different between DM and gas, leaving imprints in this transition region as probed by optical and SZ surveys. Second, the extended gas distribution beyond DM splashback must be taken into account when quantifying baryon fraction associated with DM haloes and cosmic web filaments as well as modeling galaxy quenching in the outskirts of clusters. Finally, the low density shock-heated gas between the accretion shock and the splashback radii may contribute to a significant fraction of the missing baryons and may play a role in early quenching of infalling galaxies. Investigating the differential dynamics of collisionless DM and collisional gas in the outskirts of galaxy clusters will be important for using the edges of galaxy clusters as laboratories for cosmology and astrophysics in the era of multi-wavelength cosmological surveys.

\section*{Acknowledgement}
We acknowledge Benedikt Diemer, Andrey Kravtsov, Philip Mansfield, and Xun Shi for useful comments on the draft. We used the publicly available python packages, \textsc{COLOSSUS} \citep{colossus} and \textsc{yt} \citep{yt} as part of our analysis pipeline. This work was supported in part by NSF AST-1412768 and the facilities and staff of the Yale Center for Research Computing.

\section*{Data Availability}
The {\em Omega500} simulations and the analysis code used in this paper are  available upon request. 

\bibliographystyle{mnras}
\bibliography{ms}

\begin{thebibliography}{}
\makeatletter
\relax
\def\mn@urlcharsother{\let\do\@makeother \do\$\do\&\do\#\do\^\do\_\do\%\do\~}
\def\mn@doi{\begingroup\mn@urlcharsother \@ifnextchar [ {\mn@doi@}
  {\mn@doi@[]}}
\def\mn@doi@[#1]#2{\def\@tempa{#1}\ifx\@tempa\@empty \href
  {http://dx.doi.org/#2} {doi:#2}\else \href {http://dx.doi.org/#2} {#1}\fi
  \endgroup}
\def\mn@eprint#1#2{\mn@eprint@#1:#2::\@nil}
\def\mn@eprint@arXiv#1{\href {http://arxiv.org/abs/#1} {{\tt arXiv:#1}}}
\def\mn@eprint@dblp#1{\href {http://dblp.uni-trier.de/rec/bibtex/#1.xml}
  {dblp:#1}}
\def\mn@eprint@#1:#2:#3:#4\@nil{\def\@tempa {#1}\def\@tempb {#2}\def\@tempc
  {#3}\ifx \@tempc \@empty \let \@tempc \@tempb \let \@tempb \@tempa \fi \ifx
  \@tempb \@empty \def\@tempb {arXiv}\fi \@ifundefined
  {mn@eprint@\@tempb}{\@tempb:\@tempc}{\expandafter \expandafter \csname
  mn@eprint@\@tempb\endcsname \expandafter{\@tempc}}}

\bibitem[\protect\citeauthoryear{{Adhikari}, {Dalal}  \&
  {Chamberlain}}{{Adhikari} et~al.}{2014}]{adhikari_etal2014}
{Adhikari} S.,  {Dalal} N.,   {Chamberlain} R.~T.,  2014, \mn@doi [\jcap]
  {10.1088/1475-7516/2014/11/019}, \href
  {http://adsabs.harvard.edu/abs/2014JCAP...11..019A} {11, 19}

\bibitem[\protect\citeauthoryear{{Arnaud}, {Pratt}, {Piffaretti},
  {B{\"o}hringer}, {Croston}  \& {Pointecouteau}}{{Arnaud}
  et~al.}{2010}]{arnaud_etal2010}
{Arnaud} M.,  {Pratt} G.~W.,  {Piffaretti} R.,  {B{\"o}hringer} H.,  {Croston}
  J.~H.,   {Pointecouteau} E.,  2010, \mn@doi [\aap]
  {10.1051/0004-6361/200913416}, \href
  {http://adsabs.harvard.edu/abs/2010A%26A...517A..92A} {517, A92}

\bibitem[\protect\citeauthoryear{{Aung}, {Nagai}, {Rozo}  \&
  {Garc{\'\i}a}}{{Aung} et~al.}{2021}]{Aung2020}
{Aung} H.,  {Nagai} D.,  {Rozo} E.,   {Garc{\'\i}a} R.,  2021, \mn@doi [\mnras]
  {10.1093/mnras/staa3994}, \href
  {https://ui.adsabs.harvard.edu/abs/2021MNRAS.502.1041A} {502, 1041}

\bibitem[\protect\citeauthoryear{{Baxter} et~al.,}{{Baxter}
  et~al.}{2017}]{Baxter2017}
{Baxter} E.,  et~al., 2017, \mn@doi [\apj] {10.3847/1538-4357/aa6ff0}, \href
  {https://ui.adsabs.harvard.edu/abs/2017ApJ...841...18B} {841, 18}

\bibitem[\protect\citeauthoryear{{Bertschinger}}{{Bertschinger}}{1985}]{bertschinger1985}
{Bertschinger} E.,  1985, \mn@doi [\apjs] {10.1086/191028}, \href
  {http://adsabs.harvard.edu/abs/1985ApJS...58...39B} {58, 39}

\bibitem[\protect\citeauthoryear{{Birnboim} \& {Dekel}}{{Birnboim} \&
  {Dekel}}{2003}]{Birnboim03}
{Birnboim} Y.,  {Dekel} A.,  2003, \mn@doi [\mnras]
  {10.1046/j.1365-8711.2003.06955.x}, \href
  {https://ui.adsabs.harvard.edu/abs/2003MNRAS.345..349B} {345, 349}

\bibitem[\protect\citeauthoryear{{Chang} et~al.,}{{Chang}
  et~al.}{2018}]{chang_etal18}
{Chang} C.,  et~al., 2018, \mn@doi [\apj] {10.3847/1538-4357/aad5e7}, \href
  {https://ui.adsabs.harvard.edu/abs/2018ApJ...864...83C} {864, 83}

\bibitem[\protect\citeauthoryear{{Diemer}}{{Diemer}}{2018}]{colossus}
{Diemer} B.,  2018, \mn@doi [\apjs] {10.3847/1538-4365/aaee8c}, \href
  {https://ui.adsabs.harvard.edu/abs/2018ApJS..239...35D} {239, 35}

\bibitem[\protect\citeauthoryear{{Diemer} \& {Kravtsov}}{{Diemer} \&
  {Kravtsov}}{2014}]{diemer_kravtsov2014}
{Diemer} B.,  {Kravtsov} A.~V.,  2014, \mn@doi [\apj]
  {10.1088/0004-637X/789/1/1}, \href
  {http://adsabs.harvard.edu/abs/2014ApJ...789....1D} {789, 1}

\bibitem[\protect\citeauthoryear{{Diemer}, {Mansfield}, {Kravtsov}  \&
  {More}}{{Diemer} et~al.}{2017}]{diemer_etal17}
{Diemer} B.,  {Mansfield} P.,  {Kravtsov} A.~V.,   {More} S.,  2017, \mn@doi
  [\apj] {10.3847/1538-4357/aa79ab}, \href
  {http://adsabs.harvard.edu/abs/2017ApJ...843..140D} {843, 140}

\bibitem[\protect\citeauthoryear{{G{\'o}rski}, {Hivon}, {Banday}, {Wandelt},
  {Hansen}, {Reinecke}  \& {Bartelmann}}{{G{\'o}rski} et~al.}{2005}]{Healpix}
{G{\'o}rski} K.~M.,  {Hivon} E.,  {Banday} A.~J.,  {Wandelt} B.~D.,  {Hansen}
  F.~K.,  {Reinecke} M.,   {Bartelmann} M.,  2005, \mn@doi [\apj]
  {10.1086/427976}, \href
  {https://ui.adsabs.harvard.edu/abs/2005ApJ...622..759G} {622, 759}

\bibitem[\protect\citeauthoryear{{Klypin}, {Kravtsov}, {Bullock}  \&
  {Primack}}{{Klypin} et~al.}{2001}]{klypin_etal2001}
{Klypin} A.,  {Kravtsov} A.~V.,  {Bullock} J.~S.,   {Primack} J.~R.,  2001,
  \mn@doi [\apj] {10.1086/321400}, \href
  {http://adsabs.harvard.edu/abs/2001ApJ...554..903K} {554, 903}

\bibitem[\protect\citeauthoryear{{Kravtsov}}{{Kravtsov}}{1999}]{kravtsov1999}
{Kravtsov} A.~V.,  1999, PhD thesis, New Mexico State Univ.

\bibitem[\protect\citeauthoryear{{Kravtsov}, {Klypin}  \& {Hoffman}}{{Kravtsov}
  et~al.}{2002}]{kravtsov_etal2002}
{Kravtsov} A.~V.,  {Klypin} A.,   {Hoffman} Y.,  2002, \mn@doi [\apj]
  {10.1086/340046}, \href
  {http://adsabs.harvard.edu/cgi-bin/nph-bib_query?bibcode=2002ApJ...571..563K}
  {571, 563}

\bibitem[\protect\citeauthoryear{{Lau}, {Nagai}, {Avestruz}, {Nelson}  \&
  {Vikhlinin}}{{Lau} et~al.}{2015}]{lau_etal2015}
{Lau} E.~T.,  {Nagai} D.,  {Avestruz} C.,  {Nelson} K.,   {Vikhlinin} A.,
  2015, \mn@doi [\apj] {10.1088/0004-637X/806/1/68}, \href
  {http://adsabs.harvard.edu/abs/2015ApJ...806...68L} {806, 68}

\bibitem[\protect\citeauthoryear{{Ludlow}, {Navarro}, {Springel}, {Jenkins},
  {Frenk}  \& {Helmi}}{{Ludlow} et~al.}{2009}]{Ludlow2009}
{Ludlow} A.~D.,  {Navarro} J.~F.,  {Springel} V.,  {Jenkins} A.,  {Frenk}
  C.~S.,   {Helmi} A.,  2009, \mn@doi [\apj] {10.1088/0004-637X/692/1/931},
  \href {https://ui.adsabs.harvard.edu/abs/2009ApJ...692..931L} {692, 931}

\bibitem[\protect\citeauthoryear{{Mansfield} \& {Kravtsov}}{{Mansfield} \&
  {Kravtsov}}{2020}]{Mansfield2020}
{Mansfield} P.,  {Kravtsov} A.~V.,  2020, \mn@doi [\mnras]
  {10.1093/mnras/staa430}, \href
  {https://ui.adsabs.harvard.edu/abs/2020MNRAS.493.4763M} {493, 4763}

\bibitem[\protect\citeauthoryear{{Mansfield}, {Kravtsov}  \&
  {Diemer}}{{Mansfield} et~al.}{2017}]{mansfield_etal17}
{Mansfield} P.,  {Kravtsov} A.~V.,   {Diemer} B.,  2017, \mn@doi [\apj]
  {10.3847/1538-4357/aa7047}, \href
  {http://adsabs.harvard.edu/abs/2017ApJ...841...34M} {841, 34}

\bibitem[\protect\citeauthoryear{{Miniati}, {Ryu}, {Kang}, {Jones}, {Cen}  \&
  {Ostriker}}{{Miniati} et~al.}{2000}]{Miniati2000}
{Miniati} F.,  {Ryu} D.,  {Kang} H.,  {Jones} T.~W.,  {Cen} R.,   {Ostriker}
  J.~P.,  2000, \mn@doi [\apj] {10.1086/317027}, \href
  {https://ui.adsabs.harvard.edu/abs/2000ApJ...542..608M} {542, 608}

\bibitem[\protect\citeauthoryear{{Molnar}, {Hearn}, {Haiman}, {Bryan}, {Evrard}
   \& {Lake}}{{Molnar} et~al.}{2009}]{molnar_etal2009}
{Molnar} S.~M.,  {Hearn} N.,  {Haiman} Z.,  {Bryan} G.,  {Evrard} A.~E.,
  {Lake} G.,  2009, \mn@doi [\apj] {10.1088/0004-637X/696/2/1640}, \href
  {http://adsabs.harvard.edu/abs/2009ApJ...696.1640M} {696, 1640}

\bibitem[\protect\citeauthoryear{{More}, {Diemer}  \& {Kravtsov}}{{More}
  et~al.}{2015}]{more_etal2015}
{More} S.,  {Diemer} B.,   {Kravtsov} A.~V.,  2015, \mn@doi [\apj]
  {10.1088/0004-637X/810/1/36}, \href
  {http://adsabs.harvard.edu/abs/2015ApJ...810...36M} {810, 36}

\bibitem[\protect\citeauthoryear{{More} et~al.,}{{More}
  et~al.}{2016}]{more_etal2016}
{More} S.,  et~al., 2016, \mn@doi [\apj] {10.3847/0004-637X/825/1/39}, \href
  {http://adsabs.harvard.edu/abs/2016ApJ...825...39M} {825, 39}

\bibitem[\protect\citeauthoryear{{Murata}, {Sunayama}, {Oguri}, {More},
  {Nishizawa}, {Nishimichi}  \& {Osato}}{{Murata} et~al.}{2020}]{Murata2020}
{Murata} R.,  {Sunayama} T.,  {Oguri} M.,  {More} S.,  {Nishizawa} A.~J.,
  {Nishimichi} T.,   {Osato} K.,  2020, \mn@doi [\pasj] {10.1093/pasj/psaa041},
  \href {https://ui.adsabs.harvard.edu/abs/2020PASJ...72...64M} {72, 64}

\bibitem[\protect\citeauthoryear{{Nagai}, {Kravtsov}  \& {Vikhlinin}}{{Nagai}
  et~al.}{2007}]{nagai_etal2007a}
{Nagai} D.,  {Kravtsov} A.~V.,   {Vikhlinin} A.,  2007, \mn@doi [\apj]
  {10.1086/521328}, \href {http://adsabs.harvard.edu/abs/2007ApJ...668....1N}
  {668, 1}

\bibitem[\protect\citeauthoryear{{Navarro}, {Frenk}  \& {White}}{{Navarro}
  et~al.}{1996}]{nfw1996}
{Navarro} J.~F.,  {Frenk} C.~S.,   {White} S.~D.~M.,  1996, \mn@doi [\apj]
  {10.1086/177173}, \href {http://adsabs.harvard.edu/abs/1996ApJ...462..563N}
  {462, 563}

\bibitem[\protect\citeauthoryear{{Nelson}, {Lau}, {Nagai}, {Rudd}  \&
  {Yu}}{{Nelson} et~al.}{2014}]{nelson_etal2014}
{Nelson} K.,  {Lau} E.~T.,  {Nagai} D.,  {Rudd} D.~H.,   {Yu} L.,  2014,
  \mn@doi [\apj] {10.1088/0004-637X/782/2/107}, \href
  {http://adsabs.harvard.edu/abs/2014ApJ...782..107N} {782, 107}

\bibitem[\protect\citeauthoryear{{Penna} \& {Dines}}{{Penna} \&
  {Dines}}{2007}]{Penna_Dines2007}
{Penna} M.~A.,  {Dines} K.~A.,  2007, \mn@doi [IEEE Transactions on Pattern
  Analysis and Machine Intelligence] {10.1109/TPAMI.2007.1114}, 29, 1673

\bibitem[\protect\citeauthoryear{{Prada}, {Klypin}, {Simonneau},
  {Betancort-Rijo}, {Patiri}, {Gottl{\"o}ber}  \& {Sanchez-Conde}}{{Prada}
  et~al.}{2006}]{Prada2006}
{Prada} F.,  {Klypin} A.~A.,  {Simonneau} E.,  {Betancort-Rijo} J.,  {Patiri}
  S.,  {Gottl{\"o}ber} S.,   {Sanchez-Conde} M.~A.,  2006, \mn@doi [\apj]
  {10.1086/504456}, \href
  {https://ui.adsabs.harvard.edu/abs/2006ApJ...645.1001P} {645, 1001}

\bibitem[\protect\citeauthoryear{{Rudd}, {Zentner}  \& {Kravtsov}}{{Rudd}
  et~al.}{2008}]{rudd_etal2008}
{Rudd} D.~H.,  {Zentner} A.~R.,   {Kravtsov} A.~V.,  2008, \mn@doi [\apj]
  {10.1086/523836}, \href {http://adsabs.harvard.edu/abs/2008ApJ...672...19R}
  {672, 19}

\bibitem[\protect\citeauthoryear{{Ryu}, {Kang}, {Hallman}  \& {Jones}}{{Ryu}
  et~al.}{2003}]{ryu_etal2003}
{Ryu} D.,  {Kang} H.,  {Hallman} E.,   {Jones} T.~W.,  2003, \mn@doi [\apj]
  {10.1086/376723}, \href {http://adsabs.harvard.edu/abs/2003ApJ...593..599R}
  {593, 599}

\bibitem[\protect\citeauthoryear{{Shi}}{{Shi}}{2016a}]{shi2016a}
{Shi} X.,  2016a, \mn@doi [\mnras] {10.1093/mnras/stw925}, \href
  {http://adsabs.harvard.edu/abs/2016MNRAS.459.3711S} {459, 3711}

\bibitem[\protect\citeauthoryear{{Shi}}{{Shi}}{2016b}]{shi2016b}
{Shi} X.,  2016b, \mn@doi [\mnras] {10.1093/mnras/stw1418}, \href
  {http://adsabs.harvard.edu/abs/2016MNRAS.461.1804S} {461, 1804}

\bibitem[\protect\citeauthoryear{{Shin} et~al.,}{{Shin}
  et~al.}{2019}]{Shin2019}
{Shin} T.,  et~al., 2019, \mn@doi [\mnras] {10.1093/mnras/stz1434}, \href
  {https://ui.adsabs.harvard.edu/abs/2019MNRAS.487.2900S} {487, 2900}

\bibitem[\protect\citeauthoryear{{Skillman}, {O'Shea}, {Hallman}, {Burns}  \&
  {Norman}}{{Skillman} et~al.}{2008}]{Skillman2008}
{Skillman} S.~W.,  {O'Shea} B.~W.,  {Hallman} E.~J.,  {Burns} J.~O.,   {Norman}
  M.~L.,  2008, \mn@doi [\apj] {10.1086/592496}, \href
  {https://ui.adsabs.harvard.edu/abs/2008ApJ...689.1063S} {689, 1063}

\bibitem[\protect\citeauthoryear{{Tomooka}, {Rozo}, {Wagoner}, {Aung}, {Nagai}
  \& {Safonova}}{{Tomooka} et~al.}{2020}]{Tomooka2020}
{Tomooka} P.,  {Rozo} E.,  {Wagoner} E.~L.,  {Aung} H.,  {Nagai} D.,
  {Safonova} S.,  2020, \mn@doi [\mnras] {10.1093/mnras/staa2841}, \href
  {https://ui.adsabs.harvard.edu/abs/2020MNRAS.499.1291T} {499, 1291}

\bibitem[\protect\citeauthoryear{{Turk}, {Smith}, {Oishi}, {Skory}, {Skillman},
  {Abel}  \& {Norman}}{{Turk} et~al.}{2011}]{yt}
{Turk} M.~J.,  {Smith} B.~D.,  {Oishi} J.~S.,  {Skory} S.,  {Skillman} S.~W.,
  {Abel} T.,   {Norman} M.~L.,  2011, \mn@doi [\apjs]
  {10.1088/0067-0049/192/1/9}, \href
  {https://ui.adsabs.harvard.edu/abs/2011ApJS..192....9T} {192, 9}

\bibitem[\protect\citeauthoryear{{Voit}, {Kay}  \& {Bryan}}{{Voit}
  et~al.}{2005}]{voit_etal2005}
{Voit} G.~M.,  {Kay} S.~T.,   {Bryan} G.~L.,  2005, \mn@doi [\mnras]
  {10.1111/j.1365-2966.2005.09621.x}, \href
  {http://adsabs.harvard.edu/abs/2005MNRAS.364..909V} {364, 909}

\bibitem[\protect\citeauthoryear{{Walker} et~al.,}{{Walker}
  et~al.}{2019}]{walker_etal19}
{Walker} S.,  et~al., 2019, \mn@doi [\ssr] {10.1007/s11214-018-0572-8}, \href
  {https://ui.adsabs.harvard.edu/abs/2019SSRv..215....7W} {215, 7}

\bibitem[\protect\citeauthoryear{{Wang}, {Mo}  \& {Jing}}{{Wang}
  et~al.}{2009}]{Wang2009}
{Wang} H.,  {Mo} H.~J.,   {Jing} Y.~P.,  2009, \mn@doi [\mnras]
  {10.1111/j.1365-2966.2009.14884.x}, \href
  {https://ui.adsabs.harvard.edu/abs/2009MNRAS.396.2249W} {396, 2249}

\bibitem[\protect\citeauthoryear{{Yu}, {Nelson}  \& {Nagai}}{{Yu}
  et~al.}{2015}]{yu_etal2015}
{Yu} L.,  {Nelson} K.,   {Nagai} D.,  2015, \mn@doi [\apj]
  {10.1088/0004-637X/807/1/12}, \href
  {http://adsabs.harvard.edu/abs/2015ApJ...807...12Y} {807, 12}

\bibitem[\protect\citeauthoryear{{Zhang}, {Churazov}, {Forman}  \&
  {Lyskova}}{{Zhang} et~al.}{2019}]{Zhang2019}
{Zhang} C.,  {Churazov} E.,  {Forman} W.~R.,   {Lyskova} N.,  2019, \mn@doi
  [\mnras] {10.1093/mnras/stz2135}, \href
  {https://ui.adsabs.harvard.edu/abs/2019MNRAS.488.5259Z} {488, 5259}

\bibitem[\protect\citeauthoryear{{Zhang}, {Churazov}, {Dolag}, {Forman}  \&
  {Zhuravleva}}{{Zhang} et~al.}{2020}]{Zhang2020}
{Zhang} C.,  {Churazov} E.,  {Dolag} K.,  {Forman} W.~R.,   {Zhuravleva} I.,
  2020, \mn@doi [\mnras] {10.1093/mnras/staa1013}, \href
  {https://ui.adsabs.harvard.edu/abs/2020MNRAS.494.4539Z} {494, 4539}

\bibitem[\protect\citeauthoryear{{Z{\"u}rcher} \& {More}}{{Z{\"u}rcher} \&
  {More}}{2019}]{Zurcher_More2019}
{Z{\"u}rcher} D.,  {More} S.,  2019, \mn@doi [\apj] {10.3847/1538-4357/ab08e8},
  \href {https://ui.adsabs.harvard.edu/abs/2019ApJ...874..184Z} {874, 184}

\makeatother
\end{thebibliography}

\bsp	
\label{lastpage}
\end{document}